\documentclass[oneside,twocolumn,9pt]{scrartcl}

\usepackage[english]{babel}
\usepackage[utf8]{inputenc}
\usepackage[T1]{fontenc}
\usepackage{graphicx}
\usepackage{exscale}
\usepackage{amsmath}
\usepackage{amssymb}
\usepackage[super,sort&compress,comma]{natbib} 
\usepackage[format=plain,indent=0pt,labelfont=bf]{caption}
\usepackage{pstricks}
\usepackage{authblk}
\usepackage{units}
\usepackage{siunitx}
\usepackage{braket}
\usepackage{float}

\typearea[0pt]{12}
\title{Local dielectric response in 1-propanol: $\alpha$-relaxation versus relaxation of mesoscale structures}
\author[1]{Peter Weigl}
\author[2]{Daniel Koestel}
\author[1]{Florian Pabst}
\author[1]{Jan Gabriel}
\author[2]{Thomas Walther}
\author[1]{Thomas Blochowicz}
\affil[1]{Institut für Festkörperphysik, TU Darmstadt, 64289 Darmstadt, Germany}
\affil[2]{Institut für angewandte Physik, TU Darmstadt, 64289 Darmstadt, Germany}

\date{}

\begin{document}

\twocolumn[
\begin{@twocolumnfalse}
\maketitle
\begin{abstract}
The dielectric Debye relaxation in monohydroxy alcohols has been subject of long-standing scientific interest and is presently believed to arise from the relaxation of transiently H-bonded supramolecular structures. Therefore, its manifestation might be expected to differ from a local dielectric probe as compared to the standard macroscopic dielectric experiment. In this work we present such local dielectric measurements obtained by triplet state solvation dynamics (TSD) and compare the results with macroscopic dielectric and light scattering data. In particular, with data from an improved TSD setup, a detailed quantitative comparison reveals that the Debye process does not significantly contribute to the local Stokes shift response function, while $\alpha$- and $\beta$-relaxations are clearly resolved. Furthermore, this comparison reveals that the structural relaxation has almost identical time constants and shape parameters in all three measurement techniques. Altogether our findings support the notion that the transiently bound chain structures lead to a strong cross-correlation contribution in macroscopic dielectric experiments, to which both light scattering and TSD are insensitive, the latter due to its local character and the former due to the molecular optical anisotropy being largely independent of the OH bonded suprastructures.
\end{abstract}
\end{@twocolumnfalse}]

\section{Introduction}
The relaxation behavior of hydrogen bonded liquids, especially of monohydroxy alcohols, is a long-standing topic. \cite{Debye1929a, Boehmer2014} In particular the so called Debye process observed by dielectric spectroscopy and the details of its microscopic origin have been subject of scientific debate. At present, it is widely agreed upon that the Debye peak, which is slower than the structural $\alpha$-relaxation, represents the relaxation of transient supramolecular structures. \cite{Boehmer2014} In the case of monoalcohols these structures are thought of as transient chains, which form due to H-bonding and which link the Debye process to the reorientation of an average end-to-end dipole vector. \cite{Gainaru2010} Although for a long time the Debye process in monoalcohols was believed to appear only in dielectric spectroscopy, it was recently identified in the shear mechanical response as well as in photon correlation spectroscopy. \cite{Gainaru2014, Gabriel2018a}

In general, broadband dielectric spectroscopy (BDS) and photon correlation spectroscopy (PCS) probe macroscopic quantities. While the former technique probes the collective reorientation of permanent molecular dipole moments, the latter is sensitive to the reorientation of the anisotropy tensor of the molecular polarizability. \cite{Berne2000, Gabriel2017a, Gabriel2018a} For monohydroxy alcohols, it turns out that the Debye process is usually strong or even dominant in the dielectric spectra and is far less important in PCS, where it is so far unobservable in primary alcohols while a small Debye-like contribution is reported in secondary alcohols. \cite{Gabriel2017a, Gabriel2018a} This leads to the conclusion that cross-correlations seem to be less important for the PCS spectra and the self-part of the correlation function dominates. \cite{Gabriel2018a}

Triplet state solvation dynamics is a truly local measurement technique, which can be understood as a local version of BDS. \cite{Richert2000a} In TSD a dye molecule is dissolved at low concentration in a solvent and is excited into the metastable long-lived triplet state by a UV laser pulse. \cite{Richert2000a} Due to the relaxation of the dipole moments of the surrounding solvent molecules, the phosphorescence spectrum of the dye is modified as a function of time. These changes are quantified by calculating the time-dependent Stokes shift from the spectra, revealing the local relaxation of the solvation shell. Depending on the change of the dipole moment of the dye on excitation, a local dielectric or mechanical experiment can be performed. \cite{Richert2000a, Wendt1998} Under the assumption of a continuum type dielectric the Stokes shift response function $C_\text{Stokes}(t)$ should follow the time-dependent electric modulus rather than the dielectric permittivity. \cite{Richert1995, Richert2000a} However, by contrast, experiments rather show an empirical connection of $C_\text{Stokes}(t)$ with the dielectric permittivity. \cite{Richert1992, Richert1994, Wagner1998, Sauer2014a} Thus, before tackling the problem whether a Debye contribution can be identified in $C_\text{Stokes}(t)$, it has to be clarified whether electric modulus or permittivity data are  more suitable for a comparison with TSD. Dielectric time domain measurements of modulus and permittivity will be used to clarify this problem in the following.

The main question of this paper is how the dynamics of transient supramolecular structures is reflected in the local TSD technique for the primary alcohol 1-propanol. In a previous investigation of this kind, \cite{Wendt1998} the authors concluded based on a rather limited set of data that the macroscopic dielectric response as well as the local dielectric solvation are dominated by the Debye process. However, the local TSD technique might not be very sensitive to cross correlations and therefore, similar to the results obtained in PCS, \cite{Gabriel2017a, Gabriel2018a} transient supramolecular structures only weakly contribute to the local Stokes-shift response function.

The starting point of this paper is the presentation of experimental details, in particular of the improved TSD setup that allows to access the Stokes shift response function in a broader time range than previously accessible, followed by the results obtained in BDS and TSD. Afterwards the detailed comparison of TSD, BDS and PCS results is given.

\section{Experimental section}
The dipolar phosphorescene chromophore quinoxaline (QX) was purchased from Alfa Aesar (98+\%) and used as received. The glass-forming solvent 1-propanol (99,9\%, anhydrous) was obtained from the same company. Prior to use, propanol was cleaned for at least \unit[24]{hours} with a \unit[3]{\si{\angstrom}} molecular sieve  and filtered with a \unit[200]{nm} syringe filter to reduce dust impurities. For TSD experiments the solute/solvent concentration was prepared to be \unit[2$\cdot 10^{-4}$]{mol/mol}.

The sample under investigation was filled into a rectangular quartz cell and mounted into an optical contact gas cryostat (CryoVac). The cryostat was equipped with two DT 670 A silicon diodes  as temperature sensors and can be temperature controlled from \unit[77]{K} up to \unit[320]{K} by a Lakeshore 336 temperature controller. To obtain consistent temperature values in different experimental setups, the temperature sensors were calibrated with the same PT 100 A/10 temperature sensor which was used for the temperature calibration in the BDS setups and PCS setup. \cite{Pabst2017a, Gabriel2017a, Gabriel2018a} By doing so, an absolute temperature accuracy of better than \unit[0.5]{K} was achieved. 

To excite different chromophores with UV laser pulses a laser system was set up generating \unit[355]{nm}, \unit[320]{nm} or \unit[266]{nm} laser pulses, respectively. A pulsed Nd:YAG laser (Spitlight 600 from Innolas) with \unit[10]{Hz} repetition rate and integrated pulse divider generates \unit[1064]{nm}, \unit[532]{nm} and \unit[355]{nm} pulses. After attenuation (to avoid bleaching of the dye) down to \unit[$\approx$2]{mJ} the latter can be used directly to excite the chromophore QX. A detailed description of how the other two UV laser wavelengths are generated can be found elsewhere. \cite{Talluto2016a, Weigl2018}

The phosphorescence emission is collected under 90$^{\circ}$ to the incident laser beam by a liquid light guide fiber (Newport, model 77566) and guided onto the entrance slit of a Czerny-Turner grating spectrograph (Shamrock 500i from Andor Technology) equipped with gratings of 150, 600, \unit[1800]{lines/mm}, respectively. The dispersed phosphorescence emission is registered with an Andor iStar 340T iCCD camera with integrated gate and delay generator. The effective active area of the camera is \unit[1330 x 512]{pixels} with a pixel size of \unit[13.5]{$\mu$m} and results in \unit[231.8]{nm}, \unit[55.8]{nm} and \unit[16.9]{nm} bandpasses for the three gratings. The absolute wavelength of the data acquisition optics was calibrated with a Hg(Ar) lamp.

In order to record the time-resolved phosphorescence emission spectra of the dye, there are at least two possible ways depending on the information of interest. A single laser pulse is followed and the gate width ($\Delta t$) of the camera is optimized for the first and fixed for all subsequent time points after the excitation of the dye. Thereby, the information about the intensity is preserved, but due to the decreasing emission intensity over time many accumulations are necessary to cover three (or more) decades in time. Alternatively, the time resolution $\Delta t/t$ can be optimized for each decade and each time point can be accumulated separately after excitation, thereby compensating the decreasing emission intensity and reducing the overall length of the measurement. More possible ways of measuring the time dependent Stokes shift are discussed elsewhere. \cite{Richert2000a}

\begin{figure}
\centering
\includegraphics[width=0.48\textwidth]{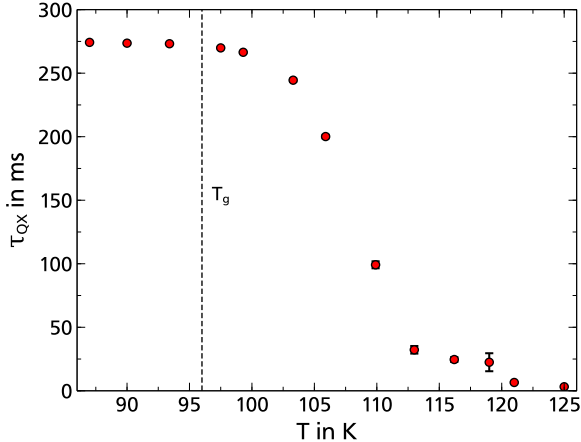}
\caption{Temperature dependence of the phosphorescence $T_{\text{1}}$ lifetime of quinoxaline $\tau_{\text{QX}}$ in 1-propanol. For $T<T_{\text{g}}$ $\tau_{\text{QX}}$ is in agreement with $\tau_{\text{QX}}$ reported at one glassy temperature in literature. \cite{Richert1990, Wagener1991, Wendt1998, Richert2000a}}\label{fig:lifetime}
\end{figure}

The lifetime measurements presented below were done in the first way because the intensity information is necessary to calculate the lifetime of a dye. 
For that purpose the gate width was set to \unit[0.1]{ms} at all temperatures and the \unit[150]{lines/mm} grating was used. The time resolved solvation measurements were all done in the second manner. Here, the time resolution $\Delta t/t$ was better than \unit[5]{\%} and the gate width was restricted to an upper limit of \unit[2]{ms} for all data points. Each spectrum consists of at least \unit[1.5$\cdot10^{6}$]{counts} at the peak maximum with the background subtracted. For all measurements the repetition time of the laser was adjusted to $\geq 3\tau_{\text{QX}}$, where $\tau_{\text{QX}}$ is the phosphorescence lifetime of quinoxaline. To calculate $\tau_{\text{QX}}$ the whole spectrum recorded at each of the time points was integrated and then fitted with an exponential decay, obtaining temperature dependent $\tau_{\text{QX}}(T)$. This result is shown in Fig.~\ref{fig:lifetime}. Starting from $T<T_{\text{g}}$, with \unit[$T_{\text{g}}=96$]{K} for 1-propanol, \cite{Takahara1994} $\tau_{\text{QX}}$ decreases by almost two orders of magnitude from \unit[$(274.2\pm4.9)$]{ms} to \unit[$(3.2\pm0.3)$]{ms} upon increasing temperature.

Such a temperature dependent phosphorescence lifetime is also observed for TSD dyes like naphthalene, quinoline and 2-naphthalenemethanolacetyl in 2-methyltetrahydrofuran. \cite{Weigl2018} While its origin is still debated, the effect in itself has been known for a long time for different phosphorescent dyes like naphthalene, benzene, toluene and several other aromatic molecules dissolved in various solvents. \cite{Tsai1968, Leubner1969, Leubner1970, Kilmer1971, Graves1972, Lin1972, Strambini1985} Some authors attributed the temperature dependence of the phosphorescence lifetime to radiative and nonradiative decays, \cite{Tsai1968, Leubner1969, Leubner1970, Kilmer1971, Graves1972, Lin1972} while other authors linked this effect to the solvents microviscosity. \cite{Strambini1985} In any case, this temperature dependence of the phosphorescence lifetime can be as much a limitation as an advantage for TSD experiments. The former due to a decreasing signal at longer delay times and the latter due to the possibility to enlarge the repetition rate of the laser $f_\text{laser}\leq1/3\tau_\text{dye}$ at higher temperatures and reduce thereby the overall measurement time.

To compare TSD data with the data of other techniques (BDS, PCS),  existing data from Ref.~\citenum{Gabriel2017a} were supplemented by measuring the electric modulus relaxation and dielectric permittivity of neat 1-propanol using a time domain dielectric setup described elsewhere in detail. \cite{Rivera2004a} Both measurements were recorded using the same sample at identical temperatures. To perform the measurement of the electric modulus relaxation a charge step is applied to realise the condition of constant dielectric displacement. Subsequently, the voltage across the sample capacitor directly reflects the variation of the eletric field in the sample. For the measurement of the dielectric permittivity the condition of constant electric field is realised by applying a voltage step across a series of sample and reference capacitors. Therefore, the voltage across the reference capacitor directly probes the polarization of the sample. Further details are described in Ref. \citenum{Rivera2004a}.

\section{Experimental results}

\subsection{Typical spectra}

\begin{figure}
\centering
\includegraphics[width=0.48\textwidth]{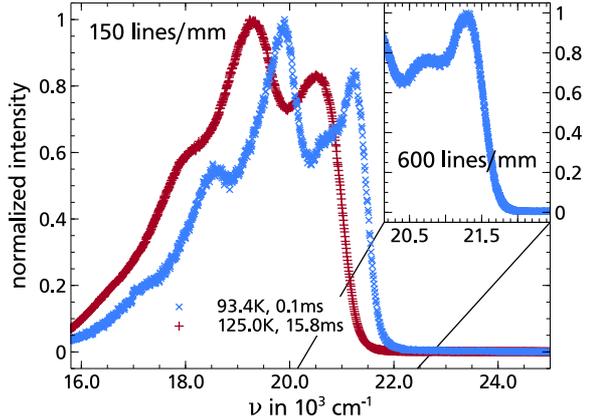}
\caption{Normalized phosphorescence emission spectra of quinoxaline in 1-propanol recorded with the \unit[150]{lines/mm} grating. Blue points refer to the unrelaxed and red points to the relaxed solvent. The absolute Stokes shift is $\Delta\nu=$\unit[(551$\pm$10)]{cm$^{-1}$}. The inset shows the spectrum for \unit[$T=93.4$]{K} recorded with the \unit[600]{lines/mm} grating showing the high energy wing with more precision. For lifetime measurements the former grating was used and for TSD experiments the latter.}\label{fig:spectra_shift}
\end{figure}

Typical phosphorescence spectra of quinoxaline in 1-propanol recorded with two different gratings are shown in Fig.~\ref{fig:spectra_shift}. For an overview, the whole spectrum was recorded with the \unit[150]{lines/mm} grating. To resolve the highest energy peak, i.e. the transition $T_{1} \rightarrow S_{0}$ $(0-0)$, with a higher spectral resolution the \unit[600]{lines/mm} grating was used.

\subsection{Triplet state solvation dynamics}
By following a common practice, \cite{Richert2000a} the high energy wing of the spectra, cf.\ inset of Fig.~\ref{fig:spectra_shift}, was fitted with a Gaussian function to obtain the mean energy $\braket{\nu}$ as a function of time and temperature. The so-called Stokes shift response function $C_\text{Stokes}(t)$ can then be calculated as:
\begin{align}
   C_\text{Stokes}(t)=\frac{\braket{\nu(t)}-\braket{\nu(\infty)}}{\braket{\nu(0)}-\braket{\nu(\infty)}},
\end{align}
where $\braket{\nu(0)}$ and $\braket{\nu(\infty)}$ are the normalization energies. The former can be determined from temperatures where the solvent is unrelaxed, e.g., the blue curve shown in Fig.~\ref{fig:spectra_shift}, while $\braket{\nu(\infty)}$ is determined at temperatures where the solvent is entirely relaxed, like from the red spectrum shown in Fig.~\ref{fig:spectra_shift}. Thus, the absolute Stokes shift $\Delta\nu=\braket{\nu(0)}-\braket{\nu(\infty)}$ can be calculated by averaging over some time points at the lowest and the highest temperature. The result of $\Delta\nu=$\unit[(551$\pm$10)]{cm$^{-1}$} is similar to results published earlier. \cite{Wendt1998}

\begin{figure}
\centering
\includegraphics[width=0.48\textwidth]{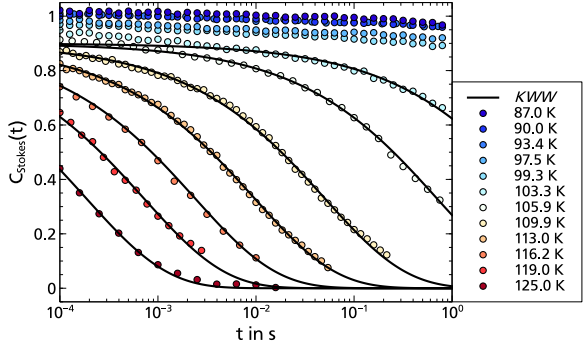}
\caption{Stokes shift response functions $C_\text{Stokes}(t)$ of quinoxaline in 1-propanol covering four decades in time. The curves are globally fitted by a \textit{KWW} function, for details see text and Fig.~\ref{fig:TSD_Master}. The data shown are in agreement with the single curve at \unit[$T=102.3$]{K} presented in. \cite{Wendt1998}} \label{fig:TSD_KWW}
\end{figure}

\begin{figure}
\centering
\includegraphics[width=0.48\textwidth]{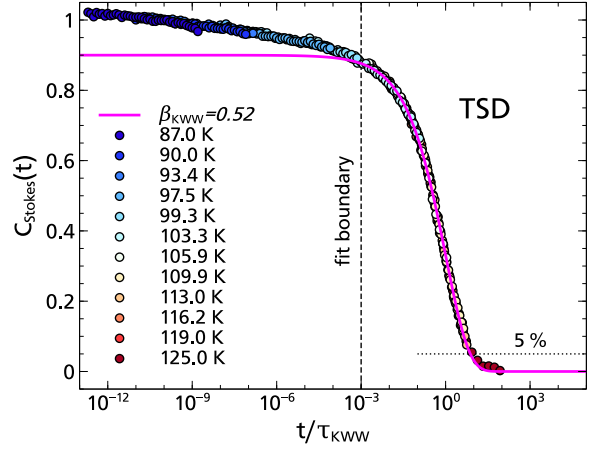}
\caption{A master plot based on the data of Fig.~\ref{fig:TSD_KWW} reveals an $\alpha$- and a $\beta$-process. But no Debye-like contribution can be distinguished down to the level of at least \unit[5]{\%}. To characterize the structural relaxation a \textit{KWW} function, i.e. $\phi(t)=\phi_{0}\cdot exp[-(t/\tau_{\text{KWW}})^{\beta_{\text{KWW}}}]$ with $\phi_{0}=0.90\pm0.01$ and $\beta_{\text{KWW}}=0.52\pm0.01$ is shown as a solid line.}\label{fig:TSD_Master}
\end{figure}

Stokes shift response functions $C_\text{Stokes}(t)$ of quinoxaline in 1-propanol are presented in Fig.~\ref{fig:TSD_KWW}, where in the temperature range from \unit[$T=87.0$]{K} up to \unit[$T=125.0$]{K} a dynamic range of four orders of magnitude is covered. Due to a decreasing $\tau_{\text{QX}}$ and thermal line broadening with increasing temperature, some curves are limited in time range for \unit[$T\geq109.9$]{K}.
To describe the data a \textit{Kohlrausch-Williams-Watts (KWW)} function was fitted globally to the structural relaxation in a master plot, cf. Fig.~\ref{fig:TSD_Master}. Therefore the \textit{KWW} functions shown in Fig.~\ref{fig:TSD_KWW} differ only in the parameter $\tau_{\text{KWW}}$. The master plot itself reveals a $\beta$-process and a structural $\alpha$-relaxation. But it does not show a Debye process separate from the structural relaxation down to the level of at least \unit[5]{\%}, cf. Fig.~\ref{fig:TSD_Master}.

\subsection{Dielectric time domain measurements}

\begin{figure}
\centering
\includegraphics[width=0.48\textwidth]{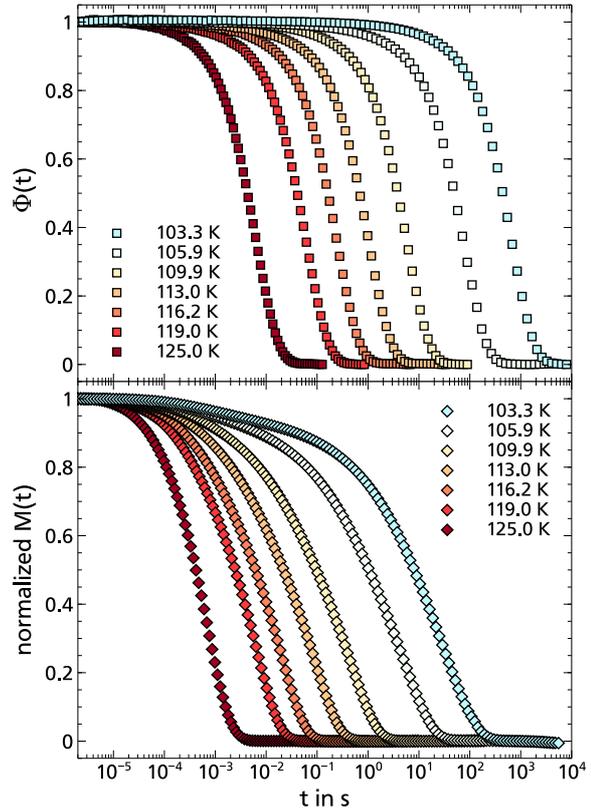}
\caption{Normalized permittivity $\Phi(t)$ (upper panel) and normalized modulus $M(t)$ (lower panel) measured in the same sample of neat 1-propanol with a time domain dielectric setup explained elsewhere in detail. \cite{Rivera2004a} The temperatures are identical to the temperatures measured in the TSD setup.}\label{fig:Rohdaten_TDD}
\end{figure}

In Fig.~\ref{fig:Rohdaten_TDD} the dielectric time domain measurements of 1-propanol are shown. It can be recognized that the different relaxation processes, like $\alpha$-, $\beta$- and Debye-relaxation, appear in a different manner in permittivity compared to electric modulus measurements. Obviously, the Debye process does not dominate the electric modulus relaxation in the same way as the dielectric permittivity and therefore it is easier to distinguish the other processes in the modulus data.

\section{Data Analysis and Discussion}
The analysis and discussion of our data will cover the line shape and the time constants of the TSD data as compared to their BDS and PCS counterparts, as well as the absolute values of the Stokes shift in the TSD data. Altogether this will provide evidence that the Debye process does not significantly contribute to the local dielectric solvation response.

Due to the broad dynamic range covered by the TSD results presented in Fig.~\ref{fig:TSD_Master}, it becomes obvious that two processes are revealed in the relaxation functions. The slower process can be described by a \textit{KWW} function with $\beta_{\text{KWW}}=0.52\pm0.01$, which is typical of the structural $\alpha$-relaxation as it is observed in other TSD experiments, e.g., for QX dissolved in 2-methyltetrahydrofuran, where the structural relaxation is described by $\beta_{\text{KWW}}=0.49$. \cite{Richert1992} As 2-methyltetrahydrofuran is a van der Waals liquid, slow relaxation contributions are not expected beyond the $\alpha$-process, in contrast to 1-propanol. So the similiar $\beta_{\text{KWW}}$ can be understood as a first indication that the Debye process, if at all,  only weakly contributes to the local Stokes shift response function in our experiment, as the latter shows the typical shape of an $\alpha$-process.

\begin{figure}[H]
\centering
\includegraphics[width=0.48\textwidth]{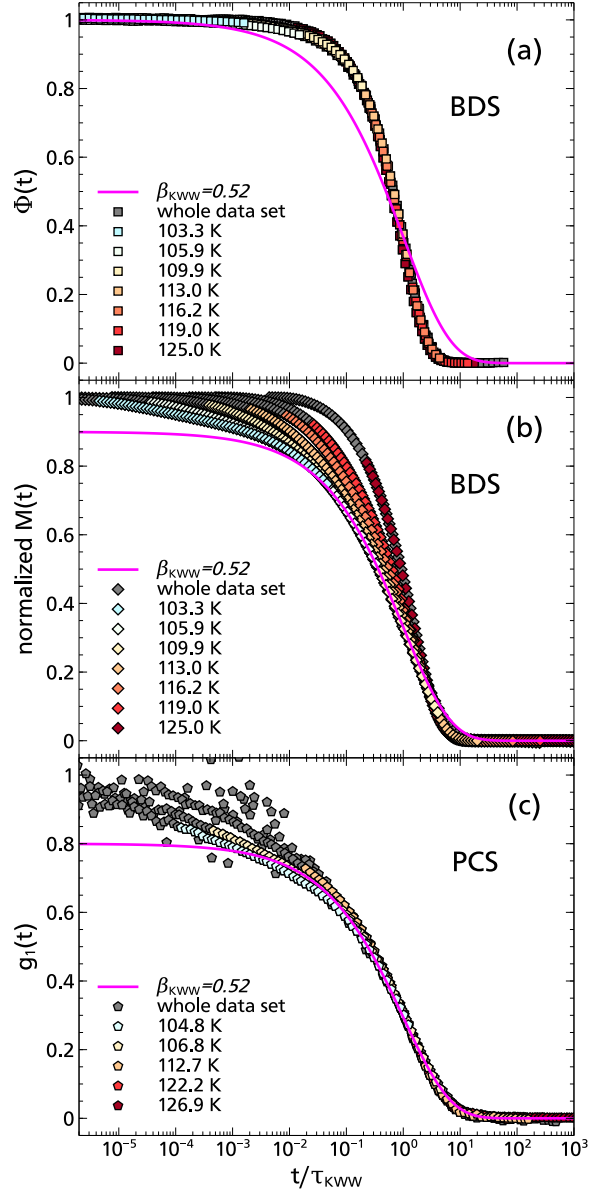}
\caption{The master plots based on the time domain dielectric data of Fig.~\ref{fig:Rohdaten_TDD}, i.e. (a): dielectric permittivity and (b): electric modulus, as well as (c): PCS data adapted from Ref.~\citenum{Gabriel2017a}. The colored points are the same data restricted to the TSD time window. The grey points are the data points from the dielectric or PCS measurements which lie beyond the TSD time window. The magenta curve is the same \textit{KWW} function as in Fig.~\ref{fig:TSD_Master}, only the amplitude is matched to $\phi_{0}=1$ (a) and to $\phi_{0}=0.8$ (c). See text for further details.}\label{fig:Masterplots}
\end{figure}

When comparing data from local dielectric solvation with macroscopic dielectric experiments, it has been a long standing issue whether $C_\text{Stokes}(t)$ should be compared to time domain modulus or permittivity data.\cite{Richert1995, Richert2000a} While standard theoretical approaches indicate that $C_\text{Stokes}$ should be close to the time-domain modulus relaxation $M(t)$, there is ample experimental evidence that  $C_\text{Stokes}(t)$ is closer to the dielectric permittivity in the time domain. \cite{Richert1992, Richert1994, Wagner1998, Sauer2014a} For the present problem this becomes particularly important as the combination of three different processes and an overall large relaxation strength leads to very different line shapes in electric modulus and permittivity representations. That will be obvious when trying to construct masterplots from the time domain data in different representations, as shown in Fig.~\ref{fig:Masterplots}. In order to be able to compare these with the master curve formed by the TSD data in Fig.~\ref{fig:TSD_Master}, all other time domain data were restricted to the same dynamic range as the TSD data, as indicated by the colored points in Fig.~\ref{fig:Masterplots}. 

All the results are shown together with the \textit{KWW} function fitted to the TSD master curve in Fig.~\ref{fig:TSD_Master} with appropriate normalization. By taking a closer look at Fig.~\ref{fig:Masterplots}b it becomes clear that time-temperature superposition fails in the case of the electric modulus because of the presence of three different relaxation processes with comparable weight ($\beta$-, $\alpha$- and Debye process) but with different temperature dependencies in the measured range. This is true even if the $M(t)$ data are restricted to the limited TSD time window, as shown by the color of the points. In the permittivity representation where the Debye process dominates, time-temperature superposition can successfully be applied but the resulting mastercurve shows a completely different shape than the one obtained from the TSD data, cf. Fig.~\ref{fig:Masterplots}a.

More insight can be gained by comparing the TSD results to PCS data. As shown in Ref.~\citenum{Gabriel2017a} for 1-propanol, the $\alpha$- and $\beta$-process in PCS and BDS are identical with respect to time constants and shape parameters within experimental accuracy, while a Debye process is not resolved with the PCS technique.\cite{Gabriel2017a} Based on the

\noindent
data of Ref.~\citenum{Gabriel2017a} a corresponding master plot was created for the PCS data as shown Fig.~\ref{fig:Masterplots}c.
Due to the much larger time range of the PCS data, failure of time temperature superposition is observed in the range of the $\beta$-relaxation while it works well in the range of the $\alpha$-process, where the shape is identical with the one obtained from the TSD data.

Thus, two conclusions can be drawn from Fig.~\ref{fig:Masterplots}: First, in agreement with a large number of observations previously reported in the literature, \cite{Richert1992, Richert1994, Wagner1998, Sauer2014a} the TSD results for dielectric solvation of the dye QX best compare with macroscopic time-domain permittivity. This cannot be seen directly from Fig.~\ref{fig:Masterplots}a, but can be concluded from the above observation that the lineshape of the PCS data, which previously were shown to be, within experimental limits, identical with the dielectric $\alpha$- and $\beta$-spectra, \cite{Gabriel2017a} matches  $C_\text{Stokes}(t)$ rather closely. Therefore, and this is the second point, in the case of 1-propanol the Debye process does not contribute significantly to the local TSD probe, very similar to the results obtained in PCS experiments.

\begin{figure}
\centering
\includegraphics[width=0.48\textwidth]{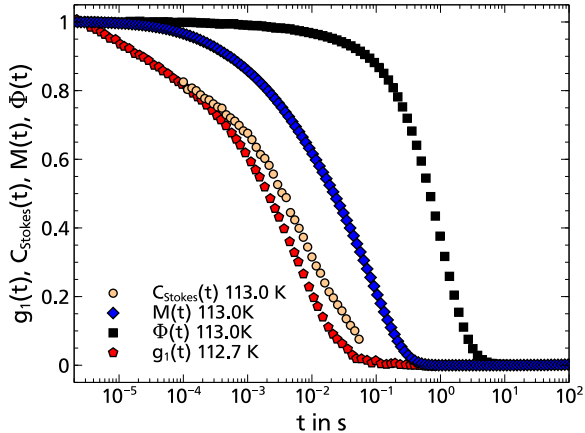}
\caption{Comparison of TSD data, i.e. $C_\text{Stokes}(t)$, with dielectric time domain data, i.e. normalized electric modulus $M(t)$ and dielectric permittivity $\Phi(t)$, and PCS data taken from Ref.~\citenum{Gabriel2017a}, i.e. $g_{1}(t)$. For this the PCS data is scaled in amplitude with a factor of $0.9/0.8$. See text for further information. }\label{fig:comp_TDD}
\end{figure}

A similar observation is made when the different response functions are compared at the same temperature, e.\,g. at $T=$\unit[113]{K}, as shown in Fig.~\ref{fig:comp_TDD}. Here it is revealed that besides the lineshape also the time constant of $C_\text{Stokes}(t)$ and $g_{1}(t)$ is almost identical. We note that in modulus representation the $\alpha$-, $\beta$-decay would be considerably faster than what is observed  in $C_\text{Stokes}(t)$, while $C_\text{Stokes}(t)$ and $g_{1}(t)$ are basically indistinguishable within the accuracy of the temperature measurements of two different experimental setups.

In order to compare time constants of TSD data with those of BDS and PCS in a more quantitative way, it is necessary to apply the same analysis to all data sets. The data presented in Ref.~\citenum{Gabriel2017a} were analyzed by using a set of well-known correlation time distributions, cf. Ref.~\citenum{Blochowicz2003}, together with the \textit{Williams-Watts approach} to connect $\alpha$- and $\beta$-process:
\begin{align}
\Phi_{\text{WWA}}(t)=\Phi_{\alpha}(t)\left((1-k)+k\,\Phi_{\beta}(t)\right),
\end{align}
where $\Phi_{\text{x}}(t)$ represents the different relaxation functions and $k$ the relative strength of the $\beta$-process. 
The $\alpha$-relaxation is characterized by the following distribution of correlation times:
\begin{align}
G_{\text{GG}}(ln\:\tau)=N_{\text{GG}}(\alpha,\beta)\:e^{-(\beta/\alpha)(\tau/\tau_{0})^{\alpha}}(\tau/\tau_{0})^{\beta},
\end{align}
while the $\beta$-process is modeled by:
\begin{align}
G_{\beta}(ln\:\tau)=N_{\beta}(a,b)\:\frac{1}{b\,(\tau/\tau_{\text{m}})^{a}+(\tau/\tau_{\text{m}})^{-ab}},
\end{align}
Details concerning the parameters are discussed elsewhere. \cite{Blochowicz2003, Gabriel2017a}

\begin{figure}
\centering
\includegraphics[width=0.48\textwidth]{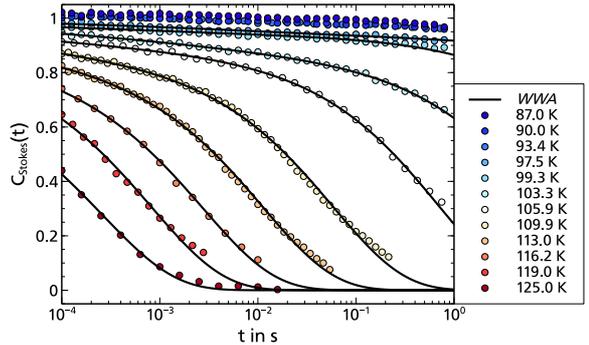}
\caption{Same TSD data as shown in Fig.~\ref{fig:TSD_KWW} but  with the \textit{Williams Watts approach (WWA)}, based on the analysis of PCS and BDS data presented in Ref.~\citenum{Gabriel2017a}, where $\alpha$- and $\beta$-process contributions are described together. See text for more information.}\label{fig:TSD_WWA}
\end{figure}

For the present purpose it suffices to say that in order to fit the TSD data, all shape parameters of $\alpha$- and $\beta$-relaxation were taken from the joint analysis of BDS and PCS data in Ref.~\citenum{Gabriel2017a}, while the strength and time constants were fitted to the TSD data. The result shows that both $\alpha$-relaxation and $\beta$-process, which is apparent in the TSD data for $T\leq$\unit[\:105.9]{K}, are well described by the fit, as shown in Fig.~\ref{fig:TSD_WWA}.

\begin{figure}
\centering
\includegraphics[width=0.48\textwidth]{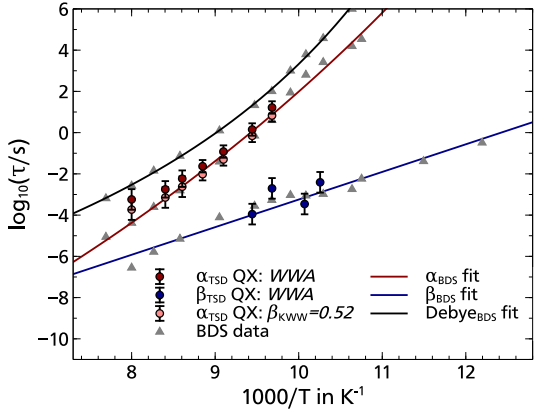}
\caption{Arrhenius plot of 1-propanol. Results from the analysis of the TSD data are shown together with results from dielectric spectroscopy (BDS) from Ref.~\citenum{Gabriel2017a}.}\label{fig:VFT}
\end{figure}

The resulting time constants are shown together with the time constants from the common \textit{KWW} analysis of the TSD data as well as with the time constants from the previous BDS data analysis, cf. Ref.~\citenum{Gabriel2017a}, in Fig.~\ref{fig:VFT}. It emerges that both the \textit{KWW} analysis and the \textit{Williams-Watts approach} analysis yield the same result for the structural $\alpha$-relaxation, in agreement with the results of BDS and PCS data. Furthermore, within the \textit{Williams Watts approach}, even the $\beta$-process of the TSD data can be evaluated at a few temperatures and again the time constants are in agreement with the previous BDS and PCS results.  Altogether, this is another indication that the Debye process does not play a significant role  for the local Stokes shift response function.

The next question to be discussed is whether the absolute value of the Stokes shift contains any indication of a significant contribution of the Debye process. First, the general observation is that there is an approximate linear relation between the absolute Stokes shift of TSD experiments and an empirical quantity called the microscopic solvent polarity ${E_{\text{T}}}^{\text{N}}$, which is approximately obeyed for QX in various polar and non-polar solvents. \cite{Wendt1998} The ${E_{\text{T}}}^{\text{N}}$ values are based on the transition energy of a dissolved betaine dye, the so called Reichardt's dye, and are normalized to tetramethylsilane (${E_{\text{T}}}^{\text{N}}=0$) and water (${E_{\text{T}}}^{\text{N}}=1$). \cite{Reichardt2010} By comparing the absolute Stokes shift of QX in 1-propanol, i.e. $\Delta\nu_{\text{1P}}=$\unit[(551$\pm$10)]{cm$^{-1}$}, with that of QX in a polar, non H-bonding, glass former like 2-methyltetrahydrofuran $\Delta\nu_{\text{MTHF}}=$\unit[250]{cm$^{-1}$}, \cite{Wendt1998} a factor $\approx 2$ between the H-bonding  and the van der Waals glass former can be noticed, while the permanent dipole moments only differ by a factor of $\approx 1.2$.\cite{Richert2000a} Accordingly the ${E_{\text{T}}}^{\text{N}}$ values are different by about a factor of three. Thus, the question arises whether the macroscopic Debye process is somehow reflected in that observed difference. However it has to be taken into consideration that both the absolute Stokes shift in the TSD experiment as well as the microscopic solvent polarity ${E_{\text{T}}}^{\text{N}}$ are local quantities, where all intermolecular forces between solvent and solute within the solvation shell play a role. Therefore, both quantities can be expected to be insensitive towards mesoscopic solvent-solvent crosscorrelations, while specific local solute-solvent interactions will play a major role. For example, it was only possible to calculate the correct  ${E_{\text{T}}}^{\text{N}}$ value of methanol, after explicitly considering a few specific solute-solvent intermolecular bindings. \cite{Alencastro1994} In particular, it is well known that the overall macroscopic relaxation strength $\Delta\epsilon$ is not a suitable measure for the microscopic solvent polarity and the absolute Stokes shift.  \cite{Reichardt2010}

To underline this fact, we compare 1-propanol with a microscopic polarity of ${E_{\text{T}}}^{\text{N}}=0.62$ with 3-phenyl-1-propanol with a microscopic polarity of ${E_{\text{T}}}^{\text{N}}=0.55$, a value close to that of 1-propanol. \cite{Reichardt2010} Although the molecular dipole moments of both molecules are the same, they largely differ in dielectric strength, as  3-phenyl-1-propanol  shows $\Delta\epsilon_{\text{3P1P}}\approx 20$, \cite{Johari1972} which is at least a factor of $\approx 4$ smaller than $\Delta\epsilon_{\text{1P}}\approx 80$ at 125K in 1-propanol, a fact, which is hardly reflected in the microscopic polarity ${E_{\text{T}}}^{\text{N}}$. In much the same way the absolute Stokes shift is unaffected by the different values of $\Delta\epsilon$, as  $\Delta\nu_{\text{1P}}=$\unit[(551$\pm$10)]{cm$^{-1}$} for 1-propanol is very close and even slightly smaller than $\Delta\nu_{\text{3P1P}}=$\unit[(587$\pm$14)]{cm$^{-1}$} of QX  in 3-phenyl-1-propanol. Obviously, the Debye process present in the macroscopic BDS of both substances in a rather different strength does not play a significant role for both, the ${E_{\text{T}}}^{\text{N}}$ value as well as the absolute TSD Stokes shift $\Delta\nu$, as expected from the above considerations.

\section{Summary and Conclusion}
By using the TSD technique over a broad dynamic range it was possible to monitor the local dielectric response in the monohydroxy alcohol 1-propanol. As compared to the macroscopic dielectric response dominated by a strong Debye process due to the relaxation of transient supramolecular chains, the local dielectric solvation response shows $\alpha$- and $\beta$-relaxation but no indication of a significant Debye contribution is observed, which is in contrast to previous conclusions for the same system. \cite{Wendt1998} A detailed comparison of the solvation response with dielectric time domain data of modulus and permittivity measurements shows that the TSD response best compares with the dielectric permittivity in accordance with several previous observations.\cite{Richert1992, Richert1994, Wagner1998, Sauer2014a} In particular, lineshape and time constants of $\alpha$- and $\beta$-relaxation are identical not only when TSD and BDS data are compared, which is more difficult due to the strong Debye contribution in BDS, but also when the recently published photon correlation results are included. \cite{Gabriel2017a} The latter comparison is much more straight forward due to the lack of a Debye contribution in the PCS data and reveals identical correlation functions within the limits of experimental accuracy.

Thus, a consistent picture emerges by a combination of data from three different experimental techniques, one of which represents a local method on a molecular scale. The interesting conclusion from this is that while $\alpha$- and $\beta$-relaxation apparently are unaffected by the locality of the technique, the Debye process presents an entirely different picture. In accordance with the idea of transient supramolecular chain structures the Debye relaxation seems to be entirely due to cross correlation terms in the macroscopic dipolar  correlation function, for which the local solvation method is obviously insensitive. However, any significant self correlation contribution on the Debye time scale that should be visible in the local TSD technique, is not detected to a substantial degree. Assuming that existence of the dye does not significantly alter the dynamic processes of its environment, this finding  implies that the dielectric Debye peak in 1-propanol is entirely due to cross correlations. Thus, a conclusion is confirmed that was already tentatively drawn in Refs.~\citenum{Gabriel2017a, Gabriel2018a} based on a comparison of dielectric and PCS data. Following this argument traces of a Debye process would be expected in the TSD data of substances like  5-methyl-2-hexanol, where PCS  already indicates some contribution of the Debye peak to the self part of the light scattering correlation function. \cite{Gabriel2018a} To clarify this point for a larger set of monohydroxy alcohols work is currently in progress.   

\section*{Conflicts of interest}
There are no conflicts to declare.

\section*{Acknowledgements}
The authors are indebted to Ranko Richert and Catalin Gainaru for stimulating and fruitful discussions. Financial support of this work by the Deutsche Forschungsgemeinschaft (DFG) in the framework of the research unit FOR 1583 under Grant No. BL1192/1 is gratefully acknowledged.

\bibliographystyle{rsc}
\bibliography{1P}
\end{document}